\documentclass{elsart}
\usepackage[dvips]{graphicx}

\begin{document}
\begin{frontmatter}
\title{Controlling soliton explosions}
\author[ivic]{J. A. Gonz\'alez\thanksref{add}},
\author[ucv]{A. Bellor\'{\i}n} and
\author[usb]{L. E. Guerrero}
\address[ivic]{Centro de F\'{\i}sica, Instituto Venezolano de Investigaciones
Cient\'{\i}ficas, Apartado 21827, Caracas 1020-A, Venezuela}
\address[ucv]{Escuela de F\'{\i}sica, Facultad de Ciencias, Universidad Central de Venezuela, Apartado 47586, Caracas 1041-A, Venezuela}
\address[usb]{Departamento de F\'{\i}sica, Universidad Sim\'on Bol\'{\i}var, Apartado Postal 89000, Caracas 1080-A, Venezuela}
\thanks[add]{Corresponding author. Fax: +58-212-5041148; e-mail: jorge@pion.ivic.ve}
\begin{keyword}
Soliton explosions; Excitable media; Control
\PACS{05.45.Yv; 03.75.Lm; 47.54.+r}
\end{keyword}
\begin{abstract}
We investigate the dynamics of solitons in generalized Klein-Gordon
equations in the presence of nonlinear damping and spatiotemporal
perturbations. We will present different mechanisms for soliton explosions.
We show (both analytically and numerically) that some space-dependent
perturbations or nonlinear damping can make the soliton internal mode
unstable leading to soliton explosion. We
will show that, in some cases, while some conditions are satisfied, the
soliton explodes becoming a permanent, extremely complex, spatiotemporal
dynamics. We believe these mechanisms can explain some of the phenomena that
recently have been reported to occur in excitable media.
We present a method for controlling soliton explosions.

\end{abstract}
\end{frontmatter}

Solitons are used in many important technological applications. Among these
applications we can mention long distance communication systems and soliton
oscillators in superconducting devices \cite{Hasegawa}.

However, under certain conditions, solitons can become unstable
\cite{Milchev,Gonzalez2,Cundiff}. Such instabilities have been called both
soliton breakup and soliton explosions. For instance, Milchev and coworkers
have studied the Frenkel-Kontorova model with anharmonic interatomic
interactions \cite{Milchev}. They have found that, in the Frenkel-Kontorova
model with nonconvex interactions between closest neighbors, a breakup of the
kink takes place when the effective amplitude of the sinusoidal substrate
potential exceeds a certain critical value.

An extensive discussion of soliton dynamics in the framework of the
Frenkel-Kontorova model can be found in the recent book \cite{Braun}.

On the other hand, in Refs. \cite{Gonzalez2} it was predicted that the soliton
internal mode can become unstable leading to soliton explosions.

Such instabilities as soliton explosions can affect all the mentioned
applications. So it is very important to understand all the possible
mechanisms of soliton explosions in order to avoid them.

In the present letter we investigate generalized Klein-Gordon equations as the
following
\begin{equation}
\phi_{tt}+R\left(  \phi_{t}\right)  -\phi_{xx}-G\left(  \phi\right)  =F(x,t),
\label{1}%
\end{equation}
where $G\left(  \phi\right)  =-\partial U\left(  \phi\right)  /\partial\phi
,$\ $U\left(  \phi\right)  $ is a potential function with at least two minima
$\phi_{1},\phi_{3}$ and a maximum $\phi_{2}$, such that $U\left(  \phi
_{1}\right)  =U\left(  \phi_{3}\right)  =0$ , $R\left(  \phi_{t}\right)  $ are
dissipative terms, and $F(x,t)$ represents external perturbations. We are
interested in kinks, that is, topological solitons between the points
$\phi_{1}$ and $\phi_{3}$. The famous sine-Gordon and $\phi^{4}$-systems are
particular cases of Eq. (\ref{1}).

The topological solitons studied in the present letter possess important
applications in condensed matter physics -- they describe domain walls in
ferromagnets and ferroelectric materials, dislocations in crystals,
charge-density waves, interphase boundaries in metal alloys, fluxons in long
Josephson junctions and Josephson transmission lines, etc. \cite{Kivshar}.

We will present different mechanisms for soliton explosions. We will show that
in some cases, while some conditions hold, the soliton explosion is permanent.

A soliton destruction is observed when inhomogeneous space-dependent
perturbations are present:
\begin{equation}
\phi_{tt}+\gamma\phi_{t}-\phi_{xx}-G\left(  \phi\right)  =F\left(  x\right)  .
\label{3}%
\end{equation}

We should say that the zeroes of $F\left(  x\right)  $ are candidates for
equilibrium positions for the soliton \cite{Gonzalez2}. If $F\left(  x\right)
$ possesses only one zero $x_{0}^{\ast}$ ($F\left(  x_{0}^{\ast}\right)  =0$),
then it is a stable position for the soliton if $\left(  \partial F/\partial
x\right)  _{x_{0}^{\ast}}>0$. Otherwise, the position is an unstable
equilibrium. The opposite is true for the antisoliton. The center of mass of a
soliton can make oscillations around a stable zero of $F\left(  x\right)  $,
and it can move away from an unstable one. However, when $F\left(  x\right)  $
has an unstable zero at $x=x_{0}^{\ast}$\ and additionally, the following
conditions holds $\lim_{x\rightarrow\pm\infty}F\left(  x\right)  =\mp
F_{\infty}$, $F_{\infty}>0$ and $F_{\infty}$ is larger than some critical
value, then the soliton can be destroyed and an antisoliton is ``created'' in
the equilibrium position $x=x_{0}^{\ast}$.

When the soliton is in an unstable equilibrium position, it is ``stretched''
by the pair of forces that are acting on its body in opposite directions. And
there is a limit for the magnitude of the pair of forces that the soliton can
resist. Nevertheless, there is a more subtle mechanism for soliton
destruction. When a soliton is close to an unstable equilibrium position, many
internal shape modes of the soliton can be excited \cite{Gonzalez1,Gonzalez2}.
If $\left(  \partial^{2}F/\partial x^{2}\right)  _{x_{0}^{\ast}}$ is larger
than some critical value, then the first internal shape mode can become
unstable and this instability can lead to the soliton destruction. In this
phenomenon the soliton decays into an antisoliton and two solitons. What is
interesting in this situation is that $F\left(  x\right)  $ can be a localized
perturbation.


Suppose we are interested in the stability of a soliton situated in
equilibrium positions created by the inhomogeneous force $F\left(  x\right)
$. Using an inverse-problem method \cite{Gonzalez2,Gonzalez3} we construct an
exact solution $\phi_{k}\left(  x\right)  $\ with the topological properties
of a kink-soliton. Then we investigate the stability of the solution solving
the spectral problem
\begin{equation}
\widehat{L}f\left(  x\right)  =\Gamma f\left(  x\right)  , \label{I}%
\end{equation}
where $\widehat{L}=-\partial_{xx}-\left[  \partial G\left(  \phi\right)
/\partial\phi\right]  _{\phi=\phi_{k}}$, $\Gamma=-\left(  \lambda^{2}%
+\gamma\lambda\right)  $.

Let us discuss some examples in detail.

The force
\begin{equation}
F\left(  x\right)  =\alpha\tanh\left(  \beta x\right)  \left[  \delta
+\varepsilon\operatorname{sech}^{2}\left(  \beta x\right)  \right]  ,
\label{II}%
\end{equation}
can sustain a kink-soliton equilibrated at point $x=0$.

When we solve the stability problem (\ref{I}) we obtain the following
eigenvalues of the discrete spectrum (for simplicity we assume $\delta=\left(
\alpha^{2}-1\right)  /2$, $\varepsilon=\left(  4\beta^{2}-\alpha^{2}\right)
/2$ and $G\left(  \phi\right)  =\left(  \phi-\phi^{3}\right)  /2$: $\Gamma
_{n}=\beta^{2}\left(  \Lambda+2\Lambda n-n^{2}\right)  -\frac{1}{2}$, where
$\Lambda\left(  \Lambda+1\right)  =3\alpha^{2}/2\beta^{2}$, $n<\Lambda$.

The translational mode of the soliton is stable if $2\beta^{2}\Lambda>1$. If
this condition is not satisfied, this just means that the soliton center of
mass will move away from the unstable equilibrium position $x=0$. This does
not necessarily lead to the soliton destruction.

However, if the following condition holds:
\begin{equation}
2\beta^{2}\left(  3\Lambda-1\right)  <1, \label{V}%
\end{equation}
the soliton first shape mode is unstable. In this case, the soliton can be destroyed!

In the very special (but also very illustrative) case $\varepsilon=0$, we have
that for $4\beta^{2}>1$, the translational mode is stable. This means the
equilibrium position $x=0$ is stable for the soliton. The soliton center of
mass can perform oscillations around point $x=0$. If $4\beta^{2}<1$ (but
$10\beta^{2}>1$), the translational mode is unstable. In this case, the
soliton can move away from point $x=0$, but it conserves its very
characteristic shape because the internal modes are still stable. However, if
$10\beta^{2}<1$, then the first internal (shape) mode becomes unstable. In
this situation the soliton can explode. If we apply a spatiotemporal
perturbation that periodically (in time) creates this instability in the place
where the soliton is situated \ at that instant, then the result will be a
highly nonstationary spatiotemporal state where the soliton is not allowed to
recover its original shape. The soliton is in a permanent explosion.

Another very important example of equation (\ref{1}) is the sine-Gordon
equation (i.e. $G(\phi)=-\sin\phi$). Suppose $F(x)=2\left(  \beta
^{2}-1\right)  \sinh\left(  \beta x\right)  \cosh^{-2}\left(  \beta x\right)
$. This perturbation creates an equilibrium position for the sine-Gordon
soliton at point $x=0$. When we solve the eigenvalue problem (\ref{I}) for the
sine-Gordon soliton in the presence of this external force we get the
following discrete spectrum: $\Gamma_{n}=\beta^{2}\left(  \Lambda+2\Lambda
n-n^{2}\right)  -1$, where $\Lambda\left(  \Lambda+1\right)  =2/\beta^{2}$.
The integer part of $\Lambda$ yields the number of eigenvalues in the discrete
spectrum. For $\beta^{2}>1$, the translational mode is stable and there are no
internal modes. If $\left(  1/3\right)  <\beta^{2}<1$, then the translational
mode is unstable (but still there are no internal modes). When $\left(
1/6\right)  <\beta^{2}<\left(  1/3\right)  $ there appears an internal shape
mode, which is stable. For $\beta^{2}<2/\left[  \Lambda_{\ast}\left(
\Lambda_{\ast}+1\right)  \right]  $, where $\Lambda_{\ast}=\left(  5+\sqrt
{17}\right)  /2$, the first internal shape mode becomes unstable. This
perturbation can destroy the sine-Gordon soliton.


We should say that a soliton, moving in a medium that is homogeneous
everywhere except for a zone where the conditions for the instabilities hold,
can undergo dramatic transient changes. But when the soliton leaves the
mentioned zone, it will return to its original steady state shape.

\begin{figure}[t]
\centerline{\includegraphics[width=3in]{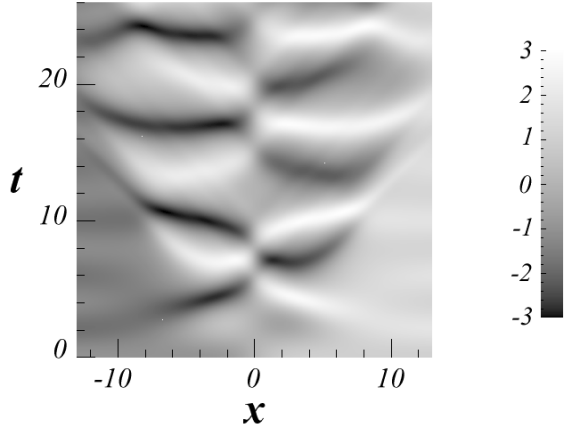}}  \caption{Initial steps
of a permanent soliton explosion sustained by a spatiotemporal perturbation as
in Eq. (\ref{4}). Here $F_{1}\left(  x\right)  =(1/2)A\left(  A^{2}-1\right)
\tanh\left(  Bx\right)  $, $F_{2}\left(  x,t\right)  =(1/2)f_{0}A\left(
4B^{2}-A^{2}\right)  \cos\left(  \omega t\right)  \sinh\left(  Bx\right)
\cosh^{-3}\left(  Bx\right)  $, $A=1.5$, $B=0.2$, $\gamma=0.1$, $\omega=1$,
$f_{0}=3$.}%
\label{fig2}%
\end{figure}

How can we produce a permanent soliton explosion? We can use time-dependent
perturbations:
\begin{equation}
\phi_{tt}+\gamma\phi_{t}-\phi_{xx}-G\left(  \phi\right)  =F_{1}\left(
x\right)  +F_{2}\left(  x,t\right)  , \label{4}%
\end{equation}
where $F_{1}\left(  x\right)  $ is a perturbation that creates a potential
well for the soliton (i.e. $F\left(  x\right)  $ possesses a zero $x_{0}%
^{\ast}$ such that $\left(  \partial F/\partial x\right)  _{x_{0}^{\ast}}>0$)
and $F_{2}\left(  x,t\right)  $ is a space-time perturbation that periodically
generates the instabilities conditions. Figure \ref{fig2} shows an example of
those highly complex spatiotemporal behaviors. In all the figures $G\left(
\phi\right)  =\left(  \phi-\phi^{3}\right)  /2$. However, similar results are
obtained with the sine-Gordon and other Generalized Klein-Gordon equations.

Can we produce permanent soliton explosions without time-dependent external perturbations?

Here we would like to remark that solitons can move with constant velocity
(without attenuation) in active and excitable media, and in systems with
nonlinear damping even without explicit external forces.

Let us discuss here briefly the importance of nonlinear damping. Linear
dissipative systems like the damped harmonic oscillator $\phi_{tt}+\gamma
\phi_{t}+\omega_{0}^{2}\phi=0$ cannot sustain oscillations. However, the
nonlinear oscillator $\phi_{tt}-b\phi_{t}+a\phi_{t}^{3}+\omega_{0}^{2}\phi=0$
supports a stable limit cycle \cite{Andronov}. The transition from a stable
focus to an unstable focus and a stable limit cycle is the result of a Hopf
bifurcation. This system is very easy to realize in practice using
negative-resistance electronic elements \cite{Chua}.

Soliton systems as the following:
\begin{equation}
\phi_{tt}+R\left(  \phi_{t}\right)  -\phi_{xx}-G\left(  \phi\right)  =0,
\label{5}%
\end{equation}
where $dR\left(  \phi_{t}\right)  /d\phi_{t}$ is negative for small values of
$\left|  \phi_{t}\right|  $ and positive elsewhere, can support solitons
moving with a constant velocity. An example of this kind of systems can be
realized in practice using a Josephson junction transmission line where the
resistor is a negative-resistance twin-tunnel-diode circuit or a
twin-transistor system \cite{Chua}. In this case, $R\left(  \phi_{t}\right)
=-b\phi_{t}+a\phi_{t}^{3}$ is a good model.

\begin{figure}[t]
\centerline{\includegraphics[width=3in]{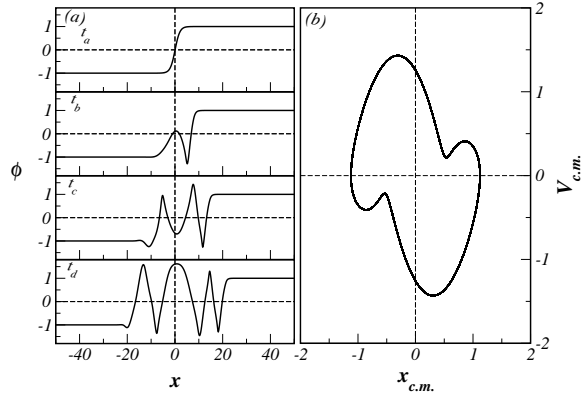}} \caption{(a) Soliton
explosion due to nonlinear damping. Here in Eq. (\ref{5}): $R\left(  \phi
_{t}\right)  =-b\phi_{t}+a\phi_{t}^{3}$, $a=1$, $b=0.7$. (b) Limit cycle
produced with the dynamics of the soliton center of mass in Eq. (\ref{8}).
Here $\Gamma\left(  x\right)  =1-l/\cosh\left(  Dx\right)  $, $F\left(
x\right)  =A\tanh\left(  Bx\right)  $, $A=0.45$, $B=0.65$, $l=2$, $D=0.65$.}%
\label{fig3}%
\end{figure}

We have investigated the shape mode stability of the soliton in the presence
of nonlinear damping as we did before using the spectral problem (\ref{I}).

Suppose $R\left(  \phi_{t}\right)  $ possesses two local extrema: a maximum
and a minimum such that the value of $\left|  R\left(  \phi_{t}\right)
\right|  $ at these extrema is $R_{m}$. If this value is comparable with the
absolute value of the extrema of $G\left(  \phi\right)  $ (let us call it
$G_{m}$), then the soliton can be destroyed. In fact, if $R_{m}>G_{m}$ the
internal shape mode of the soliton can be unstable and the soliton becomes a
highly nonstationary state.

When $F\left(  x\right)  $ in Eq. (\ref{3}) has a stable zero, say
$x=x_{0}^{\ast}$, the center of mass of a soliton can perform damped
oscillations around $x_{0}^{\ast}$. If we wish to sustain these oscillations
without explicit time-periodic external forces, then we should resort again to
negative damping. Another way to experiment negative damping is when the
damping coefficient in Eq. (\ref{3}) is a function of $x$:
\begin{equation}
\phi_{tt}+\Gamma\left(  x\right)  \phi_{t}-\phi_{xx}-G\left(  \phi\right)
=F\left(  x\right)  . \label{8}%
\end{equation}

Here $\Gamma\left(  x\right)  $ is negative in a neighborhood of $x_{0}^{\ast
}$ and positive elsewhere. This can be done in a chain of nonlinear
oscillators using negative-resistance circuits \cite{Chua} only in some small
interval of the chain. An example of $\Gamma\left(  x\right)  $ with the
required features is $\Gamma\left(  x\right)  =\gamma\left[  1-L/\cosh
^{2}\left(  Dx\right)  \right]  $, where $\left(  1-L\right)  <0$.

\begin{figure}[t]
\centerline{\includegraphics[width=3in]{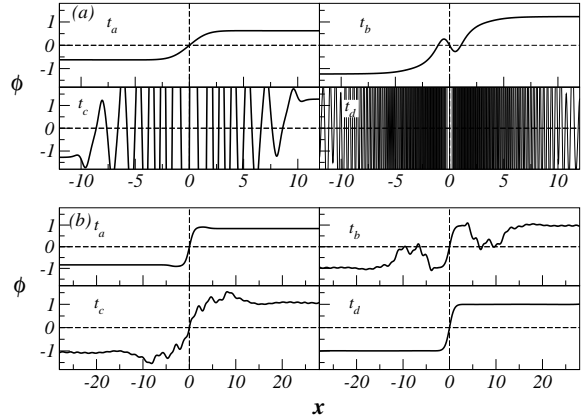}}
\caption{(a) Highly nonstationary spatiotemporal dynamics produced by
Eq. (\ref{8}). Here the simulated
equation is the same as in Fig. \ref{fig3} (b) with $l=6$. (b) The soliton dynamics can be controlled
in order to avoid the soliton explosion. Here in Eq. (\ref{9}) $\gamma=0.1$,
$F_{p}\left(  x,t\right)  = -0.385 \tanh\left(
Bx\right)  \cos\left(  \omega t\right)  $, $F_{c}\left(  x,t\right)
= -0.75  \tanh\left(  Bx\right)  \cosh^{-2}\left(
Bx\right)  $, $B=1$, $\omega=0.2$.}%
\label{fig4}%
\end{figure}

Figure \ref{fig3} (b) shows a limit cycle which is the result of the dynamics
of the soliton center of mass in Eq. (\ref{8}). However if we are not careful,
the soliton can explode also in this system. We have solved the soliton
stability problem for this equation. The most important result is that the
first internal shape mode is unstable for $L>5/2$. This behavior can be
observed in Fig. \ref{fig4} (a).\ 

Can we control all these types of explosive dynamics? Well, if we can change
the parameters of the system, then we can use parameter values that do not
lead to soliton explosions. However, sometimes we can not change the
parameters. We just are allowed to apply some external perturbation.

Let us pose the following problem:
\begin{equation}
\phi_{tt}+\gamma\phi_{t}-\phi_{xx}-G\left(  \phi\right)  =F_{p}\left(
x,t\right)  +F_{c}\left(  x\right)  . \label{9}%
\end{equation}

Equation (\ref{9}) represents a system with explosive behavior as that shown
in figures \ref{fig2}, \ref{fig3} (a) and \ref{fig4} (a), when the control
perturbation $F_{c}\left(  x\right)  =0$. The problem is to find a controlling
perturbation $F_{c}\left(  x\right)  $. Suppose $F_{p}\left(  x,t\right)  $ is
a function that periodically generates the instability conditions discussed in
the first part of the paper.\ An example can be the following $F_{p}\left(
x,t\right)  =a\cos\left(  \omega t\right)  \tanh\left(  Bx\right)  $. The
strategy could be to find a perturbation $F_{c}\left(  x\right)  $ such that
the superposition $F_{p}\left(  x,t\right)  +F_{c}\left(  x\right)  $ does not
satisfy the instability condition anymore for any $t$.

It is remarkable that this can be achieved using a localized perturbation. For
instance $F_{p}\left(  x,t\right)  =-0.385\cos\left(  \omega t\right)
\tanh\left(  Bx\right)  $ is a turbulent-producing perturbation, and
$F_{c}\left(  x\right)  =0.75\sinh\left(  Bx\right)  \cosh^{-3}\left(
Bx\right)  $ can control this behavior. This can be seen in Fig. \ref{fig4} (b).

Similarly, the turbulence created by nonlinear damping can be controlled with
a stabilizing perturbation: $\phi_{tt}-\phi_{xx}-b\phi_{t}+a\phi_{t}%
^{3}-G\left(  \phi\right)  =F_{c}\left(  x\right)  $ .

The explanation for these phenomena is based on our analytical results
presented above. The perturbation $F_{p}\left(  x,t\right)  =-0.385\cos\left(
\omega t\right)  \tanh\left(  Bx\right)  $ is able to destroy the soliton and
produce a highly nonstationary state because the perturbation
$F(x)=-0.385\tanh\left(  Bx\right)  $ leads to the instability of the soliton
shape mode (according to our condition (\ref{V})). So $F_{p}\left(
x,t\right)  $ will produce this condition regularly. The soliton will be
exposed to this instability again and again. The soliton destruction produces
several new solitons and antisolitons which also can be later destroyed
because the perturbation makes them unstable too. The control perturbation
$F_{c}\left(  x\right)  $ is able to stabilize the soliton because the total
perturbation $F\left(  x,t\right)  =-0.385\cos\left(  \omega t\right)
\tanh\left(  Bx\right)  +0.75\sinh\left(  \beta x\right)  \cosh^{-3}\left(
\beta x\right)  $ does not satisfy the shape mode instability condition for
any $t$. That is, when we solve the stability problem (\ref{I}) for
$F(x)=-0.385\mu\tanh\left(  Bx\right)  +0.75\sinh\left(  \beta x\right)
\cosh^{-3}\left(  \beta x\right)  $, the internal shape modes are always
stable for $-1\leq\mu\leq1$.

The kink-solitons are examples of a very general phenomenon called topological
defects. This set of phenomena includes: topological solitons, vortices and
spirals \cite{Hohenberg}. Although these objects can possess different origin
and nature in different physical systems, they all possess very similar
dynamical properties \cite{Hohenberg}.

The breakup of topological defects has been observed in experiments in many
systems \cite{Courtemanche}.

In different experiments, it has been observed that one topological defect can
breakup into several topological defects. In particular, the ``elementary''
breakup that we have found, where one topological defect breaks up into three
topological defects: one antidefect and two defects, has been observed in
cardiac tissue \cite{Panfilov1}.

All the situations discussed in the present paper that lead to very complex
spatiotemporal behaviors start with soliton breakups (see figures \ref{fig2},
\ref{fig3} (a), and \ref{fig4}).

At least, the following two different breakup scenarios are documented in
experiments \cite{Belmonte,Ouyang2,Barkley}. In one case, the breakup (leading
to turbulence) occurs when a spatiotemporal external forcing is added to the
system \cite{Belmonte}. In a second case, the topological defects break after
a Hopf bifurcation \cite{Barkley}.

We believe that the results of the present paper show that very similar
phenomena can occur with kink-solitons in Klein-Gordon systems. We have been
able to produce defect-mediated turbulence using spatiotemporal external
forcing, and after a Hopf bifurcation generated by nonlinear damping.

A. Bellor\'{\i}n would like to thank OPSU-CNU for support.

\end{document}